\documentclass{optica-article}

\makeatletter
\def\@articletype{}
\def\@journalname{}
\makeatother

\usepackage{lineno}
\usepackage{mathcomp}
\usepackage[subrefformat=parens]{subcaption}

\begin{document}
\pagestyle{empty}

\title{Single-shot high-resolution spectroscopy of single-photon-level optical pulses using a virtually imaged phased-array and single-photon avalanche diode array}

\author{Yuki Nagoro,\authormark{1,2} Hidehito Sato,\authormark{4} Hiroyuki Tezuka,\authormark{4,$\dag$} and Tomoyuki Horikiri\authormark{1,2,3,*}}

\address{\authormark{1}Department of Physics, Graduate School of Engineering Science, Yokohama National University, 79-5, Tokiwadai, Hodogaya, Yokohama, 240-8501, Japan\\
\authormark{2}LQUOM Inc., 79-5, Tokiwadai, Hodogaya, Yokohama, 240-8501, Japan\\
\authormark{3}Institute for Multidisciplinary Sciences, Yokohama National University, 79-5, Tokiwadai, Hodogaya, Yokohama, 240-8501, Japan\\
\authormark{4}Advanced Research Laboratory, I \& S, Digital \& Technology Platform, Sony Group Corporation, 1-7-1, Konan, Minato-ku, Tokyo, 108-0075, Japan}

\email{\authormark{*}tomoyuki.horikiri@lquom.com, \authormark{$\dag$}hiroyuki.tezuka@sony.com} 


\begin{abstract*} 
Single-shot high-resolution spectroscopy at the single-photon-level has emerged as a promising measurement technique, enabling novel observations and evaluations that were previously challenging. This technology is particularly effective for spectroscopic applications aimed at realizing frequency-multiplexed quantum repeaters. In this study, we propose a single-shot high-resolution single-photon spectroscopy system that integrates high-resolution frequency-to-spatial mode mapping using a virtually imaged phased-array (VIPA) and high-precision spatial mode detection using a single-photon avalanche diode (SPAD) array.
We experimentally demonstrated the principle of this system using weak coherent pulses with a frequency mode interval of 120$\,\mathrm{MHz}$. This interval closely matches the minimum frequency mode spacing of the atomic frequency comb quantum memory with the Pr$^{3+}$-ion-doped Y$_2$SiO$_5$ crystal.
By applying the proposed system, we expect to maximize the multiplexing capability of frequency-multiplexed quantum repeater schemes utilizing such quantum memories.
\end{abstract*}

\section{Introduction}
\label{sec:intro}
Spectroscopic techniques provide indispensable measurement tools across a wide range of scientific and technological fields. 
In particular, spectroscopic methods capable of single-photon-level measurements, high frequency (wavelength) resolution, and single-shot detection have been extensively proposed. 
However, spectroscopic systems integrating these features are scarce. 
Single-shot high-resolution spectroscopy at the single-photon-level holds the potential to enable novel observations and analysis that were previously unattainable. 
For instance, nitrogen-vacancy (NV) centers in diamond are expected to serve as highly sensitive magnetic field detectors in quantum sensing and as quantum memories in quantum communication and computation\cite{RevModPhys.89.035002, Humphreys2018}. 
Spectroscopic analysis of single photons emitted from NV centers is effective for these applications. 
Specifically, detecting subtle frequency shifts in the zero-phonon line (ZPL) can contribute to improving sensing accuracy for external environments~(i.e. electric fields) and evaluating the stability of photon sources \cite{PhysRevLett.108.206401}.
Additionally, in single-molecule fluorescence spectroscopy, capturing instantaneous spectral changes of molecular emissions in the visible light range at the single-photon level can enhance the precision of molecular identification and dynamic analysis \cite{chu2024single}. 
Thus, such advanced technology is expected to establish a novel measurement platform for spectroscopy across a broad range of applications.

Furthermore, such spectroscopy capability, enabling single-photon detection with high spectral resolution in a single-shot, is critical for realizing the quantum internet \cite{kimble2008quantum}. 
The quantum internet enables various applications, such as quantum key distribution (QKD) \cite{PhysRevLett.68.557}, distributed quantum computation \cite{PhysRevA.59.4249}, and blind quantum computation \cite{broadbent2009universal}, through entanglement distribution over long distances.
However, transmission losses over long distances present a significant challenge for entanglement distribution, and quantum repeaters \cite{PhysRevLett.81.5932} are attracting attention as a promising solution.
In quantum repeaters, quantum memories (QMs) and Bell-state measurements (BSMs) enable entanglement swapping, thereby enabling long-distance entanglement distribution. 
BSM is a measurement that projects to the Bell state, which allows entanglement swapping. 
QMs synchronize the timing of entanglement swapping and allow entanglement generation between nodes. 

Recent research has focused on more efficient and high-rate quantum repeater methods, with multiplexing in degrees of freedom such as space, time, and frequency emerging as a promising approach. 
Among these, frequency multiplexing \cite{PhysRevLett.113.053603, wengerowsky2018entanglement} is advantageous for scalability, as it minimizes switching in communication paths and allows for expansion without increasing time constraints.
Frequency-multiplexed quantum repeaters require QMs capable of frequency multiplexing. 
Absorption-based QMs with atomic frequency comb (AFC) structures, formed by comb-like absorption lines in the inhomogeneous broadening of rare-earth-doped crystals, are particularly promising for their multi-mode capacity in time and frequency domains \cite{PhysRevA.79.052329}. 
In this study, we focus on QMs based on the Pr$^{3+}$-ion-doped Y$_2$SiO$_5$ crystal (Pr:YSO). 
In the case of the AFC in Pr:YSO, the upper bandwidth limit of one frequency mode is about 4 MHz and the lower limit of the interval between frequency modes is approximately 100$\,\mathrm{MHz}$ \cite{ortu2022multimode}. 
These limits are determined by the energy level spacing of Pr:YSO.  
The region where AFCs are created is limited by the inhomogeneous broadening of Pr:YSO ($\sim$10$\,\mathrm{GHz}$).  
Therefore, in frequency-multiplexed quantum repeaters based on AFCs in Pr:YSO, the upper bound of the entanglement distribution rate can be increased by minimizing the frequency mode spacing.
As a frequency-multiplexed photon-pair source, the cavity-enhanced spontaneous parametric down conversion (cSPDC) is promising \cite{Rieländer_2016}. 
Notably, D. Lago-Rivera et al. demonstrated entanglement generation between QMs based on AFC in Pr:YSO using cSPDC as the photon-pair source \cite{Lago-Rivera2021}. 
The frequency mode spacing of photons generated by cSPDC is determined by the free spectral range (FSR) of the cavity. When a continuous-wave (CW) laser is used as the pump for cSPDC, photons in multiple frequency modes can be generated simultaneously, thereby requiring single-shot frequency mode measurements. 
Therefore, to realize frequency-multiplexed quantum repeaters using QMs based on AFCs in Pr:YSO, it is essential to perform single-shot, high-resolution identification of photon frequency modes with a spacing of about 100$\,\mathrm{MHz}$.

Conventional methods using diffraction grating or Fabry–Pérot interferometers face challenges in achieving single-shot, high-resolution spectroscopy at the single-photon level due to fundamental limitations in resolution and system.
Additionally, existing high-sensitivity detectors, such as superconducting transition-edge sensors (TESs) or superconducting single-photon detectors (SSPDs), require cryogenic operation, posing challenges in scalability and system complexity, particularly for multi-channel configurations. 
In contrast, single-photon avalanche diodes (SPADs) based on complementary metal-oxide-semiconductor (CMOS) technology offer room-temperature operation, high sensitivity, and ease of large-scale array integration, making them highly promising detection platforms. 
By integrating a virtually imaged phased-array (VIPA) with a SPAD array, we propose a single-photon spectroscopy system with a frequency resolution on the order of 100$\,\mathrm{MHz}$ and demonstrate its feasibility through proof-of-principle experiments. 
This system enables efficient identification of photon frequency modes, spatially separated via frequency-to-spatial mapping by a VIPA and directly detected by a SPAD array. 
Using weak coherent pulses with a frequency mode spacing of 120$\,\mathrm{MHz}$, we validate the effectiveness of the proposed system, which holds great promise as a spectroscopic technology for realizing frequency-multiplexed quantum repeaters and high-precision optical frequency measurements across various fields.

\section{System Configuration} 
\label{sec:system_setup}
\begin{figure}[htbp]
\centering\includegraphics[width=0.8\columnwidth]{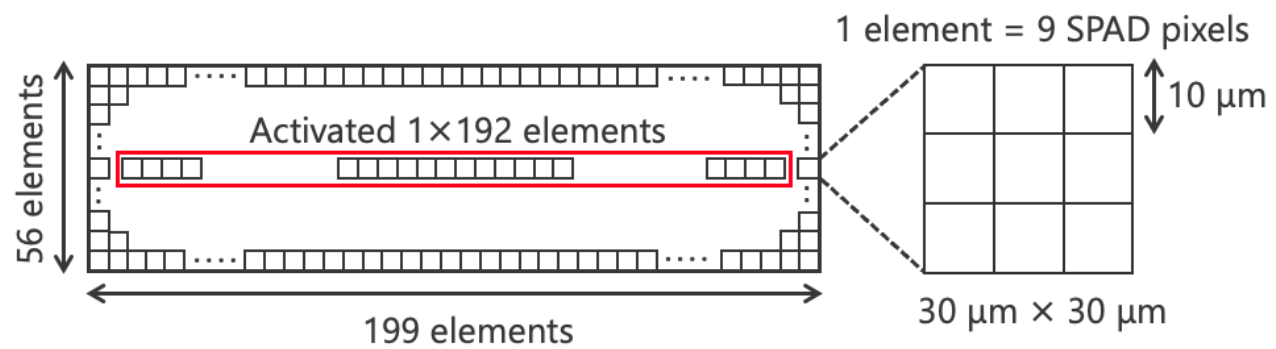}
\caption{Structure of the SPAD array.}
\label{SPAD}
\end{figure}

Single-photon avalanche diode (SPAD) pixels detect single photons by amplifying carriers generated by incident photons through avalanche multiplication. 
In this study, we utilized a commercially available SPAD array sensor (IMX560) manufactured by Sony Semiconductor Solutions Corporation. 
Each SPAD element in this array consists of 9 SPAD pixels of 10.08$\,\mu\mathrm{m}$ square, forming a square group of pixels approximately 30$\,\mu\mathrm{m}$ on each side.
As shown in Figure~\ref{SPAD}, the sensor contains a total of 56 $\times$ 199 effective elements, of which a single row of 1 $\times$ 192 elements near the center of the sensor was activated for this experiment, and the output data was recorded. 
This device is designed for low-noise measurements, ensuring that the dark count rate (DCR) does not significantly affect single-photon detection at room temperature (i.e. $\sim\,$10 counts per element per second at 25$\,^{\circ}\mathrm{C}$).

Virtually imaged phased-array (VIPA) \cite{Shirasaki:96, shirasaki1999virtually} is a type of Fabry–Pérot etalon composed of two parallel flat plates.
The output side of the plate is coated with a partially reflective film (reflectivity $r\sim95\%$), while the input side is coated with a highly reflective film ($R\sim100\%$) except for a window area, which is left uncoated.
A collimated beam is focused by a cylindrical lens into the VIPA at a small angle through this window, undergoing multiple internal reflections between the parallel plates.
This configuration produces angular dispersion through interference between the multiply reflected beams.
The output from the VIPA can be interpreted as interference between multiple virtual sources formed by these internal reflections.
The propagation angle of the output beam is determined by the phase difference between the virtual sources, which depends on the frequency of the input beam.
By focusing these beams with a lens, the VIPA functions as a frequency-to-spatial mode mapper (FSMM), converting the frequency information of light into spatial positions.

\begin{figure}[htbp]
\centering\includegraphics[width=0.9\columnwidth]{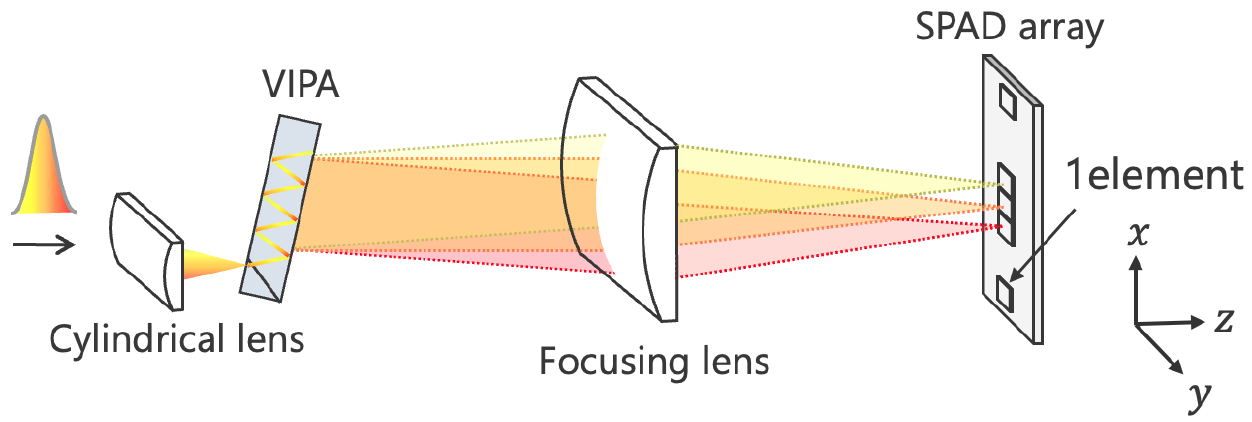}
\caption{Overview of the spectroscopic system. Light incident on the VIPA, is dispersed at frequency-dependent angles, and is spatially separated by a focusing lens. Each frequency mode is focused onto a different element of the SPAD array, enabling single-shot frequency mode identification.}
\label{system}
\end{figure}

In this study, we used two cylindrical lenses to focus beam in both the $x$-axis and $y$-axis directions, enhancing the flexibility of the optical design. 
This configuration allows efficient coupling of spatially separated beams into fibers or detectors while maintaining resolution \cite{xiao2005eight}.
Figure~\ref{system} illustrates the spectroscopic system.  
Incident light is emitted at frequency-dependent angles by the VIPA and is spatially separated by focusing lens.
The separated light is focused onto different elements of the SPAD array, where it is detected. 
This setting enables single-shot identification of frequency modes. 

In recent years, an important study by T. Chakraborty et al. reported a VIPA-based demultiplexing technique aimed at realizing frequency-multiplexed quantum repeaters~\cite{Chakraborty2025}.
In their work, idler photons generated by cSPDC with a frequency spacing of 6.5$\,\mathrm{GHz}$ were spatially mapped using a VIPA and then guided via a fiber array to SSPDs. The overall efficiency of their spectroscopic system, including VIPA transmission and fiber coupling losses, reached up to 17\%.
In contrast, our proposed system enables direct detection of photons spatially mapped by the VIPA using a SPAD array, without requiring fiber coupling.
This approach eliminates fiber coupling losses and, by utilizing SPADs that operate at room temperature, offers practical advantages in terms of ease of implementation and system scalability.

\section{Experimental Setup} 
\label{sec:experimental_setup}
To construct the high-resolution spectroscopy system using a VIPA, we designed the optical system based on the model proposed by S. Xiao et al. \cite{xiao2005eight, xiao2004dispersion}. 
This model allows the formation of arbitrary spectral patterns on the focal plane by appropriately setting the beam incident angle into the VIPA $\theta_\mathrm{in}$ and the focal lengths of the cylindrical lenses $f_\mathrm{in}, f_x,f_y$. 
Spectral dispersion is assumed to occur in the $x$-axis direction due to the VIPA.
The output intensity distribution $I_\mathrm{out}(x, y, \lambda)$ on the common focal plane of the two cylindrical lenses is expressed as
\begin{equation}
I_\mathrm{out}(x, y, \lambda) \propto \
\exp \left(-\frac{2W^2 \pi y^2}{\lambda^2 f^2_y} \right) \exp \left(-\frac{2f^2_\mathrm{in}x^2}{f^2_x W^2} \right) \
\frac{\left[1-(Rr)^{N+1}\right]^2 +4(Rr)^{N+1} \sin^2{\left[\frac{(N+1)\phi}{2} \right]}}{(1-Rr)^2 +4Rr\sin^2{\big( \frac{\phi}{2}\big)}},
\label{intensity}
\end{equation}
where
\begin{equation}
\phi = \frac{4\pi t n_r \cos{\theta_\mathrm{in}}}{\lambda} -\frac{4\pi t \tan{\theta_\mathrm{in}\cos{\theta}}}{\lambda f_x}x -\frac{2\pi t \cos{\theta_\mathrm{in}}}{n_r \lambda f^2_x}x^2,
\end{equation}
$\lambda$ is the wavelength of light, $n_r$ is the refractive index within the VIPA etalon, $t$ is the thickness of the VIPA, $W$ is the radius of the collimated beam incident to a cylindrical lens (distance at which the intensity decreases to $1/e^2$ of the maximum value), $\theta$ is the incident angle into the VIPA, and $\theta_{\mathrm{in}} = \theta/n_r$ is the internal angle inside the VIPA. 
The number of virtual light sources $N$ is given by $N = L/(2t\tan\theta_{\mathrm{in}})$, where $L$ is the length of the VIPA.

For the wavelength $\lambda_p$ at which the intensity is maximized, the following relationship holds from Eq. \eqref{intensity}:
\begin{equation}
\frac{\phi}{2} = \frac{2\pi t n_r \cos{\theta_\mathrm{in}}}{\lambda_p} -\frac{2\pi t \tan{\theta_\mathrm{in}\cos{\theta}}}{\lambda_p f_x}x -\frac{\pi t \cos{\theta_\mathrm{in}}}{n_r \lambda_p f^2_x}x^2 =m\pi.
\label{phi}
\end{equation}
Specifically, at $x=0$, the following relationship holds for the central wavelength $\lambda_0$, where the intensity is at its maximum:
\begin{equation}
m\lambda_0 = 2tn_r \cos{\theta_\mathrm{in}},
\label{m_lamda}
\end{equation}
where $m$ is an integer. 
This equation allows the determination of the appropriate incident angle $\theta_\mathrm{in}$ for the specified central wavelength $\lambda_0$.

Furthermore, from Eq. \eqref{phi} and \eqref{m_lamda}, the wavelength difference $\Delta \lambda$ is obtained as follows:
\begin{equation}
\Delta \lambda = \lambda_p -\lambda_0 =-\lambda_0\left(\frac{\tan{\theta_\mathrm{in}}\cos{\theta}}{n_r \cos{\theta_\mathrm{in}}} \frac{x}{f_x} + \frac{1}{2n^2_r} \frac{x^2}{f^2_x}  \right).
\label{delta}
\end{equation}
Based on Eq.~\eqref{delta}, it becomes possible to specify the wavelength or frequency difference corresponding to the spatial shift distance $x$ and set the focal length $f_x$ accordingly.

Additionally, the condition for the incident beam not to be clipped at the VIPA aperture is given by
\begin{equation}
t \tan \frac{\theta}{n_r} \geq w_0 \left(1 +\frac{\lambda^2 t^2}{\pi^2 w^4_0 n^2_r}  \right)^\frac{1}{2},
\label{clip_condition}
\end{equation}
where the beam waist at the input plate surface $w_0$ is expressed as $w_0 = f_{\mathrm{in}} \lambda / \pi W$, and the focal length $f_{\mathrm{in}}$ is determined from the range satisfying Eq. \eqref{clip_condition}.

In this experiment, we used a solid VIPA (LightMachinery) with an input-side reflectivity $R$ of 99.6$\%$, an output-side reflectivity $r$ of 94.5$\%$, a refractive index $n_r$ of 1.46, a thickness $t$ of 6.74$\,\mathrm{mm}$, a length $L$ of 18$\,\mathrm{mm}$, and a FSR of approximately 15$\,\mathrm{GHz}$. 

During measurement, the SPAD array was set to activate only the $1 \times 192$ elements row in the $x$-axis direction, limiting the width in the $y$-axis direction to a single element (approximately 30$\,\mu\mathrm{m}$ square). 
By restricting the $y$-axis range to the region with strong incident light, the signal-to-noise ratio (SNR) of the measurement results was improved, and the influence of ambient light noise was suppressed.

We aimed to perform single-photon-level spectroscopy of optical pulses with a frequency mode interval of 120$\,\mathrm{MHz}$. This value is close to the minimum frequency mode spacing of quantum memories based on AFCs in Pr:YSO.
To achieve high resolution, the incident angle was set to $\theta_{\mathrm{in}} \approx 0.68^\circ$ based on Eq. \eqref{m_lamda} for the central wavelength $\lambda_0$ of 605.9773$\,\mathrm{nm}$.
The focal length $f_x=1016\,\mathrm{mm}$ was derived from Eq. \eqref{delta} to obtain a spatial shift of one element (30$\,\mu\mathrm{m}$) for a frequency spacing of 120$\,\mathrm{MHz}$. 
Furthermore, by setting $W$ to 1.0$\,\mathrm{mm}$, the appropriate range of $f_{\mathrm{in}}$ derived from Eq. \eqref{clip_condition} was 57 to 415$\,\mathrm{mm}$.
Based on the condition that the beam size in the $y$-axis direction must fit within one element, $f_y$ of 40$\,\mathrm{mm}$ was adopted.

\begin{figure}[htbp]
\centering\includegraphics[width=0.9\columnwidth]{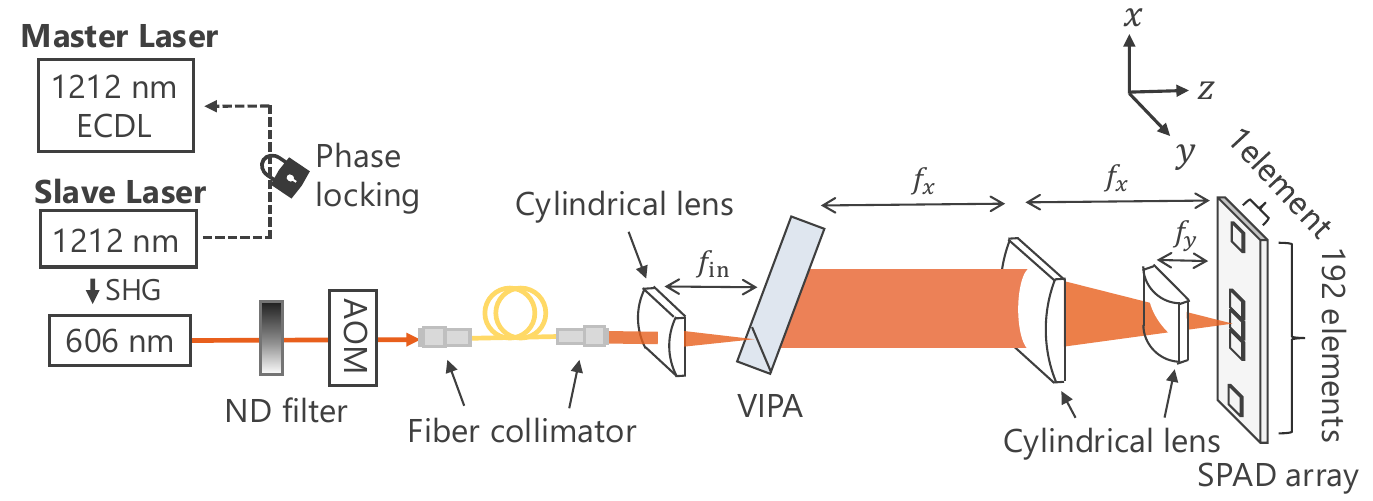}
\caption{Experimental setup. The light source used was the second harmonic generation (SHG) of an external cavity diode laser (ECDL) with a wavelength of 1212$\,\mathrm{nm}$. This CW laser was phase-locked to a master laser that was frequency-stabilized to an optical frequency comb, with a linewidth on the order of 100$\,\mathrm{kHz}$. This laser was shaped into Gaussian pulses using an AOM. The pulses were adjusted to an average photon number of 1 and incident on the SPAD array. Measurements were conducted for three frequency modes: $\lambda_0$, $\lambda_0+120\,\mathrm{MHz}$, and $\lambda_0+240\,\mathrm{MHz}$.}
\label{setup}
\end{figure}

Figure~\ref{setup} shows the experimental setup. 
The focal lengths of the lenses were selected from commercially available products closest to the calculated values, using lenses with $f_{\mathrm{in}} = 400\,\mathrm{mm}$, $f_x = 1000\,\mathrm{mm}$, and $f_y = 40\,\mathrm{mm}$. 
The light source was a CW laser (Toptica, TA-SHG pro), shaped into Gaussian pulses with a full width at half maximum (FWHM) of 180$\,\mathrm{ns}$ using an acousto-optic modulator (AOM). This pulse width is suitable for storage in the QM based on AFCs in Pr:YSO\cite{ortu2022multimode} .
The pulse intensity was adjusted with an ND filter to achieve an average photon number of one. 
The measurements were conducted for three frequency modes: the central wavelength $\lambda_0$ of 605.9773$\,\mathrm{nm}$, and frequencies modulated by +120$\,\mathrm{MHz}$ and +240$\,\mathrm{MHz}$ from $\lambda_0$. 
For each mode, 2550 pulses were directed onto the SPAD array.

\section{Results}
\label{sec:results}
\begin{figure}[htbp]
\centering\includegraphics[width=0.7\columnwidth]{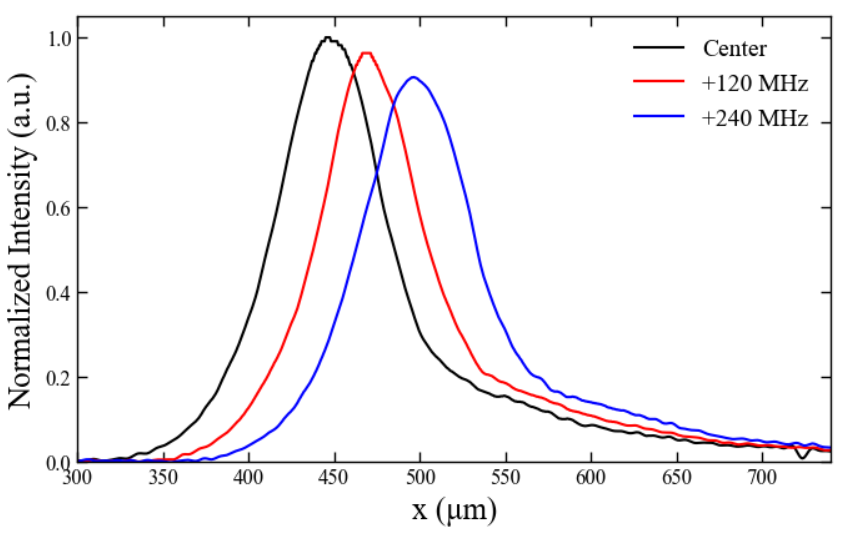}
\caption{Beam profiles along the $x$-axis under frequency modulation at 0, +120, and +240$\,\mathrm{MHz}$, measured using a beam profiler.}
\label{beamprofiler}
\end{figure}

First, to confirm that the constructed spectroscopy system operates according to theoretical design, we directly injected a CW laser instead of optical pulses and evaluated the spatial shift associated with frequency modulation.
We measured the beam position in the direction of spectral dispersion ($x$-axis) using a beam profiler (Thorlabs, BP209-VIS/M) .
Figure~\ref{beamprofiler} shows the spatial profiles observed at $\lambda_0$ and at frequencies detuned by +120$\,\mathrm{MHz}$ and +240$\,\mathrm{MHz}$.
As shown in Figure~\ref{beamprofiler}, a spatial shift of approximately one element (30$\,\mu\mathrm{m}$) was observed for a 120$\,\mathrm{MHz}$ frequency modulation, confirming that the experimental resolution matches the theoretical design.

\begin{figure}[htbp]
\centering\includegraphics[width=1.0\columnwidth]{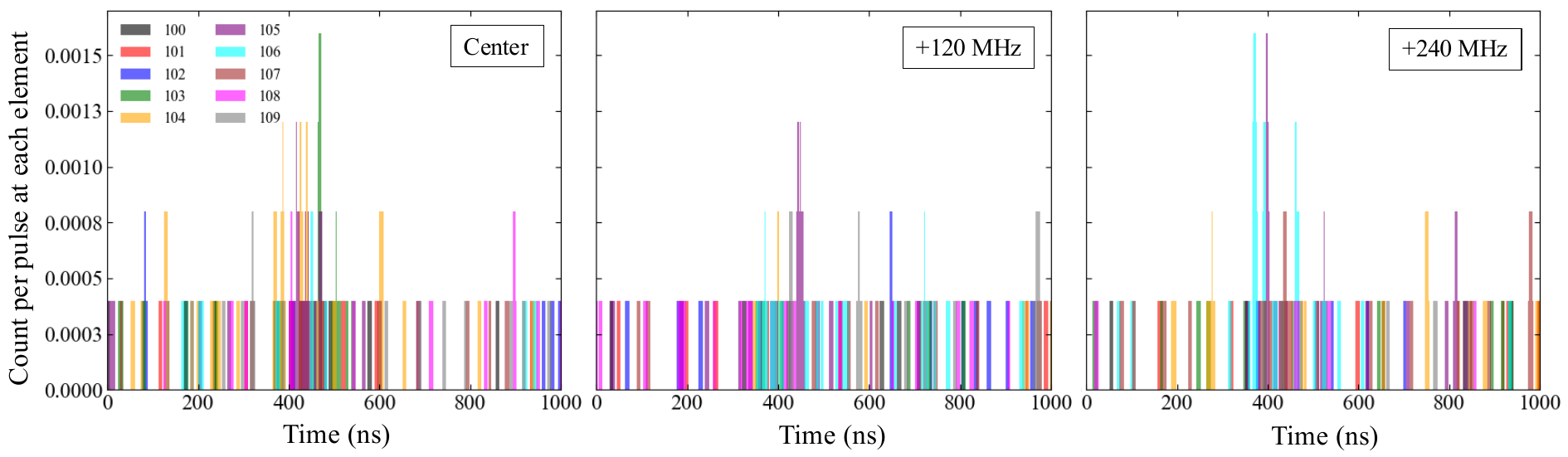}
\caption{Time histograms of photon detection counts for elements 100 to 109. Gaussian pulses with an average photon number of 1 were input into the system for three frequency modes. The detection counts for each element were normalized to the number of incident pulses (2550) and plotted as the average count per pulse. }
\label{time}
\end{figure}

Next, we conducted experiments to detect Gaussian pulses with an average photon number of one using the SPAD array.
The optical axis was aligned so that beam was incident near element number 100 of the activated 192 elements of the SPAD array.
Photon detection was performed with a time resolution of 1$\,\mathrm{ns}$ for three frequency modes ($\lambda_0$, $\lambda_0+120\,\mathrm{MHz}$, $\lambda_0+240\,\mathrm{MHz}$). 
The detection counts for each element were normalized by the number of incident pulses (2550) and calculated as the average count per pulse. Figure~\ref{time} shows the time distribution for elements 100 to 109.

As shown in Figure~\ref{time}, the peak of the detection counts shifted to different elements with frequency modulation. 
The peak is observed at element 103 for $\lambda_0$, at element 105 for $\lambda_0+120\,\mathrm{MHz}$, and at elements 105–106 for $\lambda_0+240\,\mathrm{MHz}$, corresponding to an approximate spatial shift of one element per modulation step.

\begin{figure}[htbp]
\centering\includegraphics[width=0.7\columnwidth]{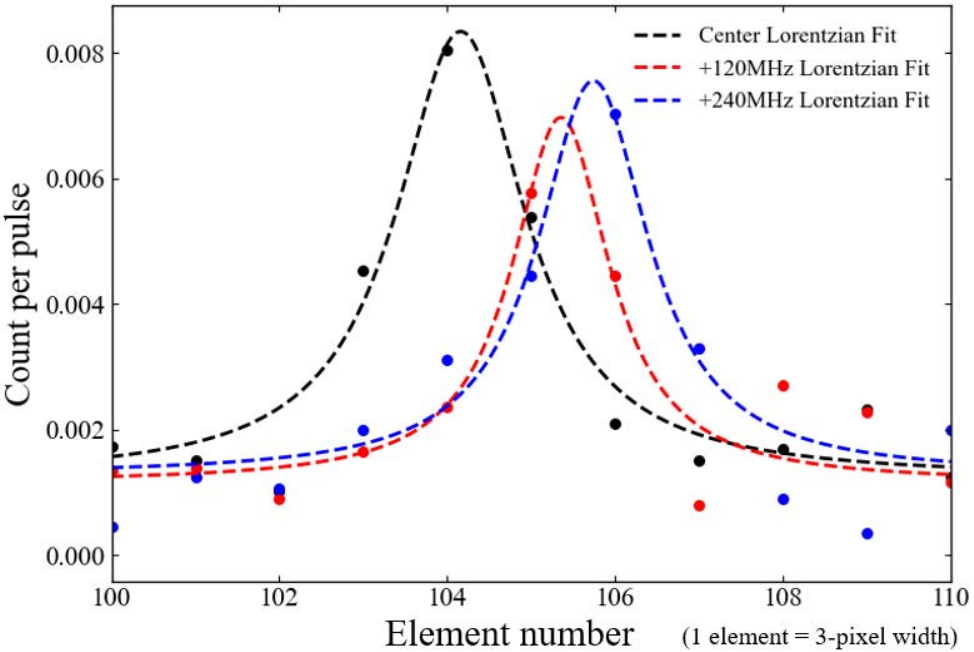}
\caption{Average detection counts per pulse for each element and Lorentzian fitting results for each frequency mode. The total detection counts within the time window of 200–650$\,\mathrm{ns}$ were integrated and normalized to obtain the average count per pulse.
The spatial distributions were fitted with Lorentzian functions.
}
\label{pixelnumber}
\end{figure}

To quantitatively evaluate the spatial separation, we calculated the total detection counts for each element within the time window corresponding to the presence of Gaussian pulses (200-650$\,\mathrm{ns}$) and normalized the results to the average count per pulse. 
This time window was determined based on histograms obtained using Gaussian pulses with a sufficiently high average photon number.
The spatial distributions were fitted with Lorentzian functions, as the output profile of a VIPA is approximated by a Lorentzian distribution \cite{xiao2005eight}.
The fitting results are shown in Figure~\ref{pixelnumber}.

Figure~\ref{pixelnumber} shows a spatial shift of approximately one element between $\lambda_0$ and $\lambda_0+120\,\mathrm{MHz}$ modulation. 
However, the shift for $\lambda_0+240\,\mathrm{MHz}$ modulation is limited to about 0.5 elements. 
This discrepancy is attributed to the beam width in the $x$-axis exceeding the width of a single element. 
Theoretically, an ideal spatial shift is expected, as illustrated in Figure~\ref{beamprofiler}. 
In this measurement, the average number of photons was kept close to one, which may have increased the variation in the photon detection positions due to the spatial spread of the beam. 
If the number of measurements is insufficient, the spatial distribution of the detection position reflecting the beam width will not converge sufficiently, leading to estimation errors in the observed spatial shift.
In the present measurement, the number of input pulses was limited to 2550 to ensure the stability of the system.
In future experiments, improving system stability is expected to allow for a larger number of measurements, thereby enhancing statistical reliability and improving the accuracy of spatial shift estimation.

\section{Discussion}
In this section, we discuss the effectiveness of introducing the proposed spectroscopic system into a frequency-multiplexed quantum repeater scheme. 
Specifically, we evaluated the effect of improving the heralding rate to examine whether this system can realistically function as a multi-mode frequency identifier.

\begin{figure}[htbp]
\centering\includegraphics[width=1.0\columnwidth]{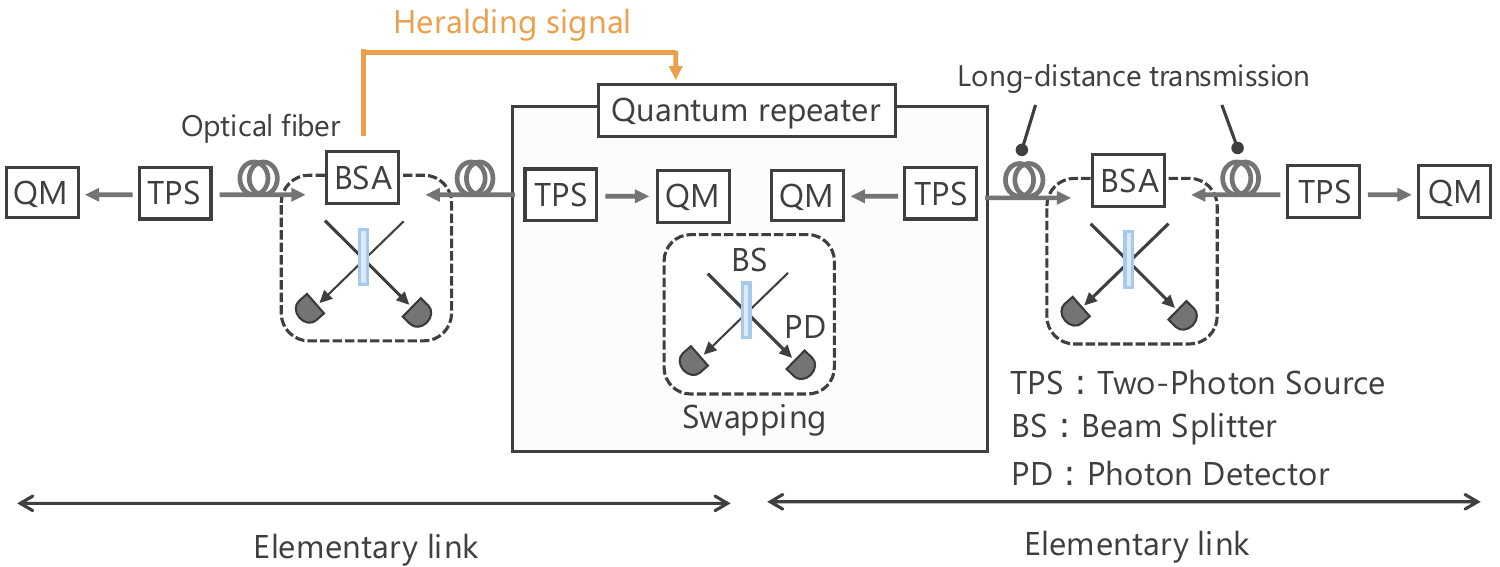}
\caption{Overview of the quantum repeater scheme. A success signal of the BSM at the BSA is a heralding signal, which enables entanglement swapping using QMs at the quantum repeater.
In this study, we assume that the BSMs outside the repeater are multiplexed. The storage time of the QM is fixed according to the reference \cite{PhysRevLett.113.053603}.}
\label{scheme}
\end{figure}

First, we describe the assumed frequency-multiplexed quantum repeater scheme and the heralding rate. 
In this paper, we focus on the scheme proposed by N. Sinclair et al. \cite{PhysRevLett.113.053603}. 
Figure~\ref{scheme} shows an overview of this scheme. 
This scheme is based on the meet-in-the-middle protocol \cite{jones2016design}, where a Bell-state analyzer (BSA) is placed at the midpoint of the communication channel. 
A detection event at the BSA serves as the heralding signal for successful entanglement generation via BSM.
This heralding signal enables entanglement swapping using the QMs at the quantum repeater. 
By frequency multiplexing this BSM, the entanglement generation rate can be significantly increased. 
Our proposed spectroscopic system enables the identification of photon frequency modes in frequency-multiplexed BSM.

This quantum repeater scheme is compatible with entanglement generation protocols based on either two-photon or one-photon interference \cite{PhysRevLett.98.190503, duan2001long}.
The two-photon interference method is strongly affected by transmission loss, while the single-photon interference method, although requiring phase stabilization, is more robust against transmission loss as it is based on single-photon detection events.  
In recent years, high-precision phase stabilization techniques have been developed. 
For example, twin-field QKD\cite{Lucamarini2018}, which also requires phase stabilization, has been demonstrated over a distance of 600 km using a technique known as dual-band phase stabilization \cite{pittaluga2021600}. 
Furthermore, the quantum repeater scheme based on single-photon interference with mild phase stabilization requirements has also been proposed \cite{yoshida2024multiplexed}. 
In this work, we chose the single-photon interference method to improve the overall repeater system efficiency.

Next, we evaluated the heralding rate based on single-photon interference. 
Assuming single-photon entanglement generation based on Fock state basis, the heralding rate is calculated as the probability of a successful BSM, defined as an event in which only one of the two detectors detects a photon within an elementary link.
Let the probability of photon generation be $p$, the detector efficiency be $\eta_\mathrm{det}$, the transmission loss coefficient be $\alpha$, and the length of elementary link be $L$. 
The success probability of the BSM in a single-mode configuration without multiplexing is expressed as follows \cite{PhysRevA.101.042301}:
\begin{equation}
P_\mathrm{single} =2p\eta_\mathrm{det} 10^{- \frac{\alpha L/2}{10}}.
\label{single}
\end{equation}
On the other hand, when $M$ modes are multiplexed, the probability that at least one mode results in a successful BSM is given by:
\begin{equation}
P_\mathrm{multi} =1-\left( 1 -2p\eta_\mathrm{VIPA} \eta_\mathrm{det} \eta_\mathrm{WC} 10^{- \frac{\alpha L/2}{10}} \right)^M,
\label{multi}
\end{equation}
where, $\eta_{\mathrm{VIPA}}$ represents the transmission efficiency of the VIPA, and $\eta_{\mathrm{WC}}$ represents the efficiency of wavelength conversion from the telecommunication band to a wavelength detectable by the SPAD. 
The heralding rate is obtained by dividing these success probabilities by the communication trial time. 
By comparing these rates, we can estimate the number of modes required for a multi-mode configuration to outperform the single-mode configuration.

\begin{figure}[htbp]
\centering\includegraphics[width=0.7\columnwidth]{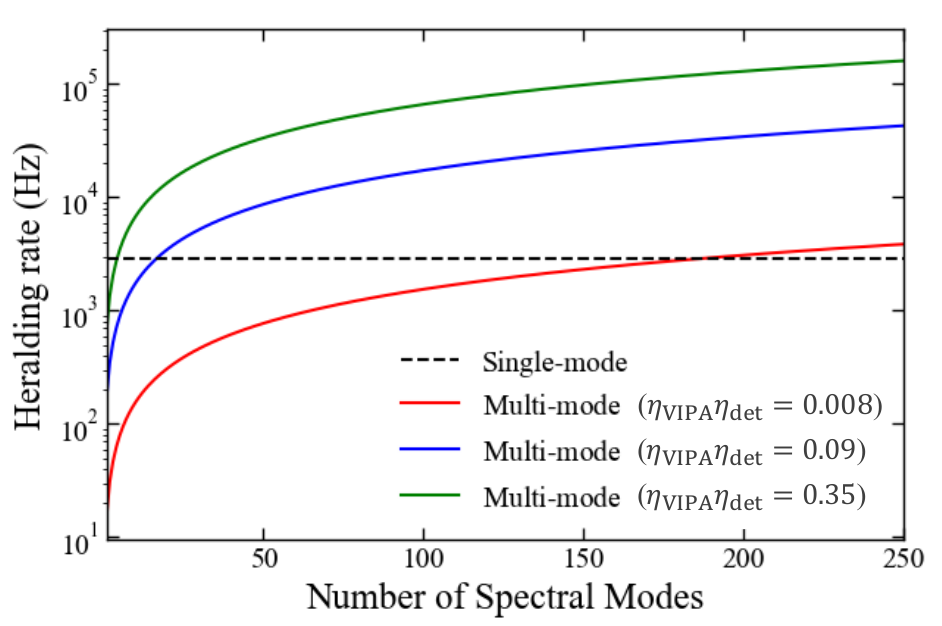}
\caption{Comparison of heralding rates in single-mode and multi-mode configurations. The extent to which frequency multiplexing improves heralding performance was evaluated in three different scenarios.}
\label{Heraldingrate}
\end{figure}

Using the experimental results for detection efficiency, we calculated the heralding rate based on Eq.~\eqref{single} and~\eqref{multi}. 
The common parameters were set as follows: $p = 0.01$, $\alpha = 0.2$, and $L = 100\,\mathrm{km}$. 
The detection efficiency in the single-mode configuration is assumed to be $\eta_{\mathrm{det}}  = 0.9$, corresponding to the performance of SSPD. 
For the multi-mode configuration, the detection efficiency is estimated to be $\eta_{\mathrm{VIPA}}\eta_{\mathrm{det}} = 0.008$ based on the experimental results shown in Figure~\ref{pixelnumber}. 
Although exact detection efficiency of a SPAD array in the 600$\,\mathrm{nm}$ band is not evaluated, the efficiency is estimated as approximately 50$\%$ or higher based on the data sheet of the SPAD array.
In this experiment, the transmission efficiency of the VIPA was measured to be approximately 69$\%$, leading to a theoretical detection efficiency under ideal conditions (i.e., when each frequency mode is fully focused within the spatial extent of a single element, without significant spreading) of $\eta_{\mathrm{VIPA}}\eta_{\mathrm{det}} = 0.35$. 
However, in the current experimental setup, the beam spread along the $x$-axis exceeds the width of a single element, reducing the practical efficiency to approximately $\eta_{\mathrm{VIPA}}\eta_{\mathrm{det}} \approx 0.09$. 
Furthermore, assuming wavelength conversion efficiency $\eta_{\mathrm{WC}} = 0.6$ \cite{niizeki2020two}, the following four scenarios are considered for calculating the heralding rate:
\begin{enumerate}
    \item Single-mode (baseline).
    \item Multi-mode ($\eta_{\mathrm{VIPA}} \eta_{\mathrm{det}} = 0.008$).
    \item Multi-mode ($\eta_{\mathrm{VIPA}} \eta_{\mathrm{det}} = 0.09$).
    \item Multi-mode ($\eta_{\mathrm{VIPA}} \eta_{\mathrm{det}} = 0.35$).
\end{enumerate}
The heralding rates obtained under these conditions are shown in Figure~\ref{Heraldingrate}.

As shown in Figure~\ref{Heraldingrate}, Scenario 2 achieves a heralding rate that surpasses the single-mode configuration when 190 or more modes are used, while Scenario 3 requires only 18 modes to do so. 
Considering the theoretical storage capacity of 100 modes \cite{ortu2022multimode} and the experimental demonstration of simultaneous storage of 15 modes in QM based on AFCs in Pr:YSO \cite{PhysRevLett.123.080502}, the proposed system demonstrates realistic scalability. Furthermore, under ideal conditions, only 6 modes are required to exceed the heralding rate of the single-mode configuration, indicating significant potential for improving communication efficiency. Although an assumed SSPD efficiency of 0.9 is close to ideal, it is reasonable to expect a reduction in the single-mode heralding rate due to coupling losses into the fiber connected to SSPD. Taking this into account, it is possible that the multi-mode heralding rate surpasses that of the single-mode case with a smaller degree of multiplexing than the estimated values here.

The experimental detection efficiency of 0.008, which is lower than the theoretical value of 0.09 for the current setup, is primarily attributed to misalignment between the optical axis and the SPAD array. 
Pre-measurements using a beam profiler confirmed that the beam width and shape were as designed, and no major issues were observed in beam control within the optical system. 
However, the beam width and its alignment with the pixel plane of the SPAD array could not be directly verified, suggesting that the pixel plane may have been offset from the ideal focal plane. 
This misalignment likely caused some photons to spread into adjacent elements or regions with lower sensitivity, thereby reducing the overall detection efficiency. 
An effective countermeasure to this issue would be to use a higher-precision multi-axis stage to adjust the position of the detection surface while injecting a CW laser with constant power and observing the counts from the SPAD array in real time. 
This is expected to quantitatively correct the deviation from the focus and improve the detection efficiency.

In future experiments, achieving the theoretical efficiency of 35$\%$ will require reducing the beam width in the $x$-axis direction to the size of one element. 
According to the results in Figure~\ref{beamprofiler}, the FWHM of the beam is approximately 70$\,\mu\mathrm{m}$. 
This spatial broadening increases crosstalk to neighboring elements, which not only reduces the detection efficiency but also limits the accuracy of frequency mode identification in single-shot measurements. 
In the current experimental setup, the beam width along the $y$-axis is designed to fall within one element. 
However, since the beam width along the x-axis is in a trade-off relationship with frequency resolution, it is challenging from an optical design perspective to suffciently narrow the beam width while maintaining a separation distance that corresponds to the frequency mode spacing. 
The $x$-axis beam width primarily depends on the focal length $f_x$ of the cylindrical lens, shortening $f_x$ reduces the beam width but also decreases the frequency resolution. 
Therefore, to suppress spatial crosstalk while maintaining adequate resolution, it is essential to employ a high-resolution VIPA element that provides sufficient spatial separation even at short focal lengths.

The frequency resolution of the VIPA can be evaluated using the FWHM of the spectrum, expressed as follows \cite{xiao2005eight, PhysRevLett.123.080502}:
\begin{equation}
\mathrm{FWHM}_\mathrm{freq} = \frac{c}{2\pi n_r t \cos{\theta}} \frac{1-Rr}{\sqrt{Rr}},
\label{FWHM}
\end{equation}
where, $c$ is the speed of light, the FSR of the VIPA is ${c}/{(2n_r t \cos{\theta})}$, and the finesse is $\pi\sqrt{Rr}/(1-Rr)$. 
This FWHM represents the theoretical minimum -3 dB bandwidth in the VIPA transmission spectrum. It corresponds to the minimum frequency mode spacing required to suppress crosstalk below -3 dB. 
The parameters of the VIPA used in this experiment are reflectivities $R=99.6\%$ and $r=94.5\%$, thickness $t=6.74\,\mathrm{mm}$, a refractive index $n_r=1.46$, and incident angle $\theta\approx1.0^\circ$, for which $\cos{\theta}\approx1$ can be assumed. 
Substituting these values into Eq.~\eqref{FWHM}, the theoretical frequency resolution is calculated as $\mathrm{FWHM}_\mathrm{freq} \approx 294$ MHz.
Therefore, a frequency mode spacing of at least approximately 300$\,\mathrm{MHz}$ is required to suppress crosstalk below -3$\,\mathrm{dB}$, making single-shot frequency mode identification at a spacing of 120$\,\mathrm{MHz}$ challenging.

\begin{figure}[htbp]
\centering\includegraphics[width=1.0\columnwidth]{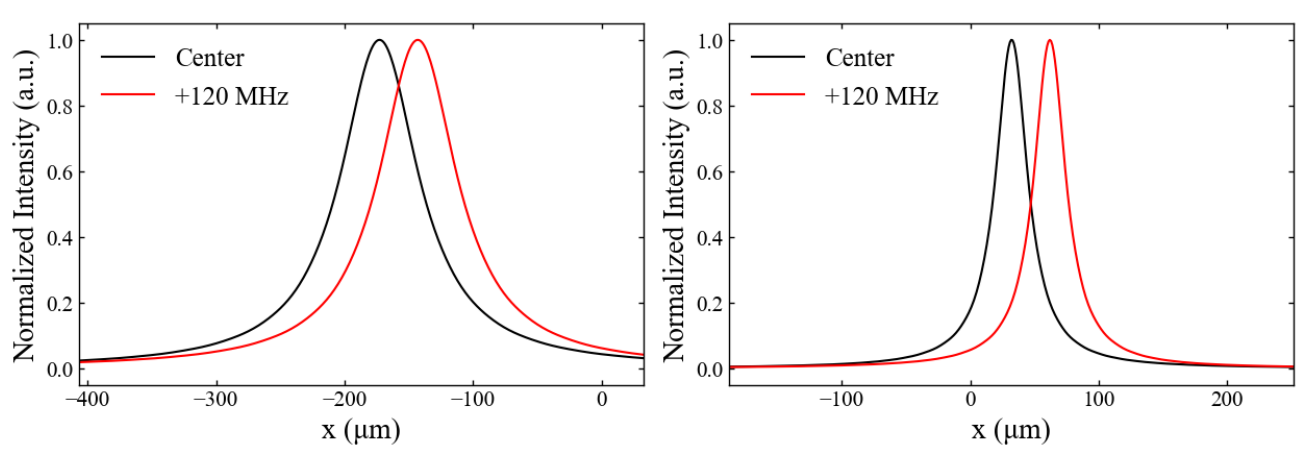}
\caption{Simulated beam profiles for the current (left) and the redesigned (right) spectroscopic system, where the VIPA element thickness was increased. The redesigned configuration reduces spatial crosstalk and enhances frequency resolution.}
\label{simulation}
\end{figure}

The $\mathrm{FWHM}_\mathrm{freq}$ is inversely proportional to the FSR of the VIPA~\cite{Xiao:05}, and increasing the thickness $t$ of the VIPA improves frequency resolution. 
By inverting Eq.~\eqref{FWHM}, the thickness required to achieve $\mathrm{FWHM}_\mathrm{freq} \approx 120\,\mathrm{MHz}$ is found to be approximately 16.5$\,\mathrm{mm}$. 
Based on this condition, the VIPA thickness was set to $t=16.5\,\mathrm{mm}$ and the spectroscopic system was redesigned accordingly. 
The parameters of the redesigned spectrometer system were $\theta_{\mathrm{in}} \approx 0.30^\circ$, $f_\mathrm{in}=498\,\mathrm{mm}$, and $f_x=449\,\mathrm{mm}$. 
We simulated the beam profile along the $x$-axis using Eq.~\eqref{intensity} under both the redesigned condition and the current setup ($\theta_{\mathrm{in}} \approx 0.68^\circ$, $f_\mathrm{in}=400\,\mathrm{mm}$, $f_x=1000\,\mathrm{mm}$). 
The resulting beam profiles are shown in Figure~\ref{simulation}.

As shown in Figure~\ref{simulation}, the redesigned spectroscopic system significantly suppresses spatial beam spread, reducing crosstalk and improving frequency resolution. 
In addition, further improvements in resolution can be achieved by increasing the reflectivity of the VIPA, which enhances its finesse~\cite{Xiao:05}. 
These results demonstrate that optimizing the structural parameters of the VIPA element can reduce crosstalk between adjacent frequency modes and enable high-precision mode identification in single-shot spectroscopy.
Introducing a high-resolution VIPA element optimized for this system is expected to realize single-shot, high-efficiency, high-resolution spectral spectroscopy.

\section{Conclusion}
\label{sec:conclusion}
In this study, we proposed a spectroscopic system integrating a VIPA and SPAD array, and demonstrated its principle using weak coherent pulses with a frequency spacing of 120$\,\mathrm{MHz}$.
This system enables single-shot, high-resolution identification of photons in different frequency modes by mapping frequency to spatial position via the VIPA and detecting them with the spatially resolved SPAD array.
Furthermore, by evaluating heralding rates, we showed that the theoretical detection efficiency of the current setup allows for multiplexing of 18 or more modes to outperform a single-mode configuration.
Considering that the entanglement generation rate between elementary links increases proportionally to the heralding rate, this advantage becomes even more pronounced in large-scale quantum repeater networks with multiple repeater nodes. 
These results indicate that this system is a promising and practical spectroscopic method for frequency-multiplexed quantum repeaters and is expected to contribute to the construction of future quantum networks. 
Additionally, further improvements in system efficiency will enable applications beyond quantum communication, extending to a wide range of photonic technologies.

\begin{backmatter}

\bmsection{Funding}
Ministry of Internal Affairs and Communications (JPMI00316); New Energy and Industrial Technology Development Organization; Moonshot Research and Development Program (JPMJMS226C).

\bmsection{Acknowledgment}
Funding listed above was provided to Y. Nagoro and T. Horikiri.
The authors thank Sony Semiconductor Solutions Corporation for their support.
Y. Nagoro acknowledges Daisuke Yoshida and Feng-Lei Hong for their technical assistance during the experiments, and Akira Ozawa for valuable discussions.

\bmsection{Disclosures}
The authors declare no conflicts of interest.

\bmsection{Data Availability Statement}
Data underlying the results presented in this paper are not publicly available at this time but may be obtained from the authors upon reasonable request.


\end{backmatter}




\bibliography{sample}

\begin{thebibliography}{10}
\newcommand{\enquote}[1]{``#1''}

\bibitem{RevModPhys.89.035002}
C.~L. Degen, F.~Reinhard, and P.~Cappellaro, \enquote{Quantum sensing,} {\protect\JournalTitle{Rev. Mod. Phys.}} \textbf{89}, 035002 (2017).

\bibitem{Humphreys2018}
P.~C. Humphreys, N.~Kalb, J.~P.~J. Morits, \emph{et~al.}, \enquote{Deterministic delivery of remote entanglement on a quantum network,} {\protect\JournalTitle{Nature}} \textbf{558}, 268--273 (2018).

\bibitem{PhysRevLett.108.206401}
V.~M. Acosta, C.~Santori, A.~Faraon, \emph{et~al.}, \enquote{Dynamic stabilization of the optical resonances of single nitrogen-vacancy centers in diamond,} {\protect\JournalTitle{Phys. Rev. Lett.}} \textbf{108}, 206401 (2012).

\bibitem{chu2024single}
J.~Chu, A.~Ejaz, K.~M. Lin, \emph{et~al.}, \enquote{Single-molecule fluorescence multiplexing by multi-parameter spectroscopic detection of nanostructured fret labels,} {\protect\JournalTitle{Nature nanotechnology}} \textbf{19}, 1150--1157 (2024).

\bibitem{kimble2008quantum}
H.~J. Kimble, \enquote{The quantum internet,} {\protect\JournalTitle{Nature}} \textbf{453}, 1023--1030 (2008).

\bibitem{PhysRevLett.68.557}
C.~H. Bennett, G.~Brassard, and N.~D. Mermin, \enquote{Quantum cryptography without bell's theorem,} {\protect\JournalTitle{Phys. Rev. Lett.}} \textbf{68}, 557--559 (1992).

\bibitem{PhysRevA.59.4249}
J.~I. Cirac, A.~K. Ekert, S.~F. Huelga, and C.~Macchiavello, \enquote{Distributed quantum computation over noisy channels,} {\protect\JournalTitle{Phys. Rev. A}} \textbf{59}, 4249--4254 (1999).

\bibitem{broadbent2009universal}
A.~Broadbent, J.~Fitzsimons, and E.~Kashefi, \enquote{Universal blind quantum computation,} in \emph{2009 50th annual IEEE symposium on foundations of computer science,}  (IEEE, 2009), pp. 517--526.

\bibitem{PhysRevLett.81.5932}
H.-J. Briegel, W.~D\"ur, J.~I. Cirac, and P.~Zoller, \enquote{Quantum repeaters: The role of imperfect local operations in quantum communication,} {\protect\JournalTitle{Phys. Rev. Lett.}} \textbf{81}, 5932--5935 (1998).

\bibitem{PhysRevLett.113.053603}
N.~Sinclair, E.~Saglamyurek, H.~Mallahzadeh, \emph{et~al.}, \enquote{Spectral multiplexing for scalable quantum photonics using an atomic frequency comb quantum memory and feed-forward control,} {\protect\JournalTitle{Phys. Rev. Lett.}} \textbf{113}, 053603 (2014).

\bibitem{wengerowsky2018entanglement}
S.~Wengerowsky, S.~K. Joshi, F.~Steinlechner, \emph{et~al.}, \enquote{An entanglement-based wavelength-multiplexed quantum communication network,} {\protect\JournalTitle{Nature}} \textbf{564}, 225--228 (2018).

\bibitem{PhysRevA.79.052329}
M.~Afzelius, C.~Simon, H.~de~Riedmatten, and N.~Gisin, \enquote{Multimode quantum memory based on atomic frequency combs,} {\protect\JournalTitle{Phys. Rev. A}} \textbf{79}, 052329 (2009).

\bibitem{ortu2022multimode}
A.~Ortu, J.~V. Rakonjac, A.~Holz{\"a}pfel, \emph{et~al.}, \enquote{Multimode capacity of atomic-frequency comb quantum memories,} {\protect\JournalTitle{Quantum Science and Technology}} \textbf{7}, 035024 (2022).

\bibitem{Rieländer_2016}
D.~Rieländer, A.~Lenhard, M.~Mazzera, and H.~d. Riedmatten, \enquote{Cavity enhanced telecom heralded single photons for spin-wave solid state quantum memories,} {\protect\JournalTitle{New Journal of Physics}} \textbf{18}, 123013 (2016).

\bibitem{Lago-Rivera2021}
D.~Lago-Rivera, S.~Grandi, J.~V. Rakonjac, \emph{et~al.}, \enquote{Telecom-heralded entanglement between multimode solid-state quantum memories,} {\protect\JournalTitle{Nature}} \textbf{594}, 37--40 (2021).

\bibitem{Shirasaki:96}
M.~Shirasaki, \enquote{Large angular dispersion by a virtually imaged phased array and its application to a wavelength demultiplexer,} {\protect\JournalTitle{Opt. Lett.}} \textbf{21}, 366--368 (1996).

\bibitem{shirasaki1999virtually}
M.~Shirasaki, A.~Akhter, and C.~Lin, \enquote{Virtually imaged phased array with graded reflectivity,} {\protect\JournalTitle{IEEE Photonics Technology Letters}} \textbf{11}, 1443--1445 (1999).

\bibitem{xiao2005eight}
S.~Xiao and A.~M. Weiner, \enquote{An eight-channel hyperfine wavelength demultiplexer using a virtually imaged phased-array (vipa),} {\protect\JournalTitle{IEEE photonics technology letters}} \textbf{17}, 372--374 (2005).

\bibitem{Chakraborty2025}
T.~Chakraborty, A.~Das, H.~van Brug, \emph{et~al.}, \enquote{Towards a spectrally multiplexed quantum repeater,} {\protect\JournalTitle{npj Quantum Information}} \textbf{11}, 3 (2025).

\bibitem{xiao2004dispersion}
S.~Xiao, A.~M. Weiner, and C.~Lin, \enquote{A dispersion law for virtually imaged phased-array spectral dispersers based on paraxial wave theory,} {\protect\JournalTitle{IEEE journal of quantum electronics}} \textbf{40}, 420--426 (2004).

\bibitem{jones2016design}
C.~Jones, D.~Kim, M.~T. Rakher, \emph{et~al.}, \enquote{Design and analysis of communication protocols for quantum repeater networks,} {\protect\JournalTitle{New Journal of Physics}} \textbf{18}, 083015 (2016).

\bibitem{PhysRevLett.98.190503}
C.~Simon, H.~de~Riedmatten, M.~Afzelius, \emph{et~al.}, \enquote{Quantum repeaters with photon pair sources and multimode memories,} {\protect\JournalTitle{Phys. Rev. Lett.}} \textbf{98}, 190503 (2007).

\bibitem{duan2001long}
L.-M. Duan, M.~D. Lukin, J.~I. Cirac, and P.~Zoller, \enquote{Long-distance quantum communication with atomic ensembles and linear optics,} {\protect\JournalTitle{Nature}} \textbf{414}, 413--418 (2001).

\bibitem{Lucamarini2018}
M.~Lucamarini, Z.-Q. Yin, Y.~Zhao, and Z.~Yuan, \enquote{Overcoming the rate-distance limit of quantum key distribution without quantum repeaters,} {\protect\JournalTitle{Nature}} \textbf{557}, 400--403 (2018).

\bibitem{pittaluga2021600}
M.~Pittaluga, M.~Minder, M.~Lucamarini, \emph{et~al.}, \enquote{600-km repeater-like quantum communications with dual-band stabilization,} {\protect\JournalTitle{Nature Photonics}} \textbf{15}, 530--535 (2021).

\bibitem{yoshida2024multiplexed}
D.~Yoshida and T.~Horikiri, \enquote{Multiplexed quantum repeaters based on single-photon interference with mild stabilization,} {\protect\JournalTitle{Communications Physics}} \textbf{7}, 367 (2024).

\bibitem{PhysRevA.101.042301}
Y.~Wu, J.~Liu, and C.~Simon, \enquote{Near-term performance of quantum repeaters with imperfect ensemble-based quantum memories,} {\protect\JournalTitle{Phys. Rev. A}} \textbf{101}, 042301 (2020).

\bibitem{niizeki2020two}
K.~Niizeki, D.~Yoshida, K.~Ito, \emph{et~al.}, \enquote{Two-photon comb with wavelength conversion and 20-km distribution for quantum communication,} {\protect\JournalTitle{Communications Physics}} \textbf{3}, 138 (2020).

\bibitem{PhysRevLett.123.080502}
A.~Seri, D.~Lago-Rivera, A.~Lenhard, \emph{et~al.}, \enquote{Quantum storage of frequency-multiplexed heralded single photons,} {\protect\JournalTitle{Phys. Rev. Lett.}} \textbf{123}, 080502 (2019).

\bibitem{Xiao:05}
S.~Xiao, A.~M. Weiner, and C.~Lin, \enquote{Experimental and theoretical study of hyperfine wdm demultiplexer performance using the virtually imaged phased-array (vipa),} {\protect\JournalTitle{J. Lightwave Technol.}} \textbf{23}, 1456 (2005).

\end{thebibliography}

\end{document}